\input{aipcheck}

\documentclass[
    ,final            % use final for the camera ready runs
  ]
  {aipproc}

\layoutstyle{8x11single}

\usepackage{amsmath}

\newcommand{\lsim}{\raise.3ex\hbox{$<$\kern-.75em\lower1ex\hbox{$\sim$}}}
\newcommand{\ima}{{\mbox{Im}\,}}
\newcommand{\rea}{{\mbox{Re}\,}}

\begin{document}

\title{Unitarized Chiral Perturbation Theory and the meson spectrum}


\author{Jos\'e R. Pel\'aez}{
  address={Dept. de F\'{\i}sica Te\'orica II. Universidad Complutense. 28040 Madrid. Spain}
}

\author{C. Hanhart}{
  address={Institut f\"ur Kernphysik and J\"ulich Center for Hadron
        Physics, Forschungzentrum J\"ulich GmbH}
}

\author{J. Nebreda}{
  address={Dept. de F\'{\i}sica Te\'orica II. Universidad Complutense. 28040 Madrid. Spain}
}

\author{G. R\'ios}{
  address={Dept. de F\'{\i}sica Te\'orica II. Universidad Complutense. 28040 Madrid. Spain}
}

\begin{abstract}
In this talk we briefly review how the unitarization of
Chiral Perturbation Theory with dispersion relations 
can successfully describe the meson-meson scattering data and generate 
light resonances, whose mass, width and nature can be related to QCD
parameters like quark masses and the number of colors.
\end{abstract}
\maketitle
\vspace*{-.2cm}

\vspace*{-.1cm}
\section{Introduction}
\vspace*{-.1cm}

Light hadron spectroscopy lies beyond the applicability regime of
perturbative QCD. However, there is a rigorous and systematic expansion in the
form of an effective field theory of QCD, 
known as Chiral Perturbation Theory (ChPT)
\cite{chpt1}, which provides a model independent description of
the dynamics of the lightest mesons, namely, the 
Goldstone Bosons of the QCD spontaneous chiral symmetry breaking. 
Despite pure ChPT is limited to low energies and masses,
here we review how, when combined with model independent
dispersion relations,  it leads to a successful description of meson dynamics, 
generating resonant states without a priori assumptions on their existence or nature.
This `` unitarized ChPT'' is a useful tool to identify
the spectroscopic nature of resonances through
their dependence on the QCD number of colors $N_c$, 
but also to relate lattice results to physical 
resonances by studying their quark mass dependence.

We will concentrate on the meson sector, where ChPT is most developed
and converges somewhat better, and
 is built out of pion, kaon and eta fields only, 
as a low energy expansion 
of a Lagrangian respecting all QCD symmetries.
Generically, it is organized in powers of
 $O(p^2/\Lambda^2)$, where $p$ stands either for derivatives, momenta or meson masses,
and $\Lambda\equiv 4 \pi f_\pi$, where
$f_\pi$ denotes the pion decay constant.
ChPT is renormalized order by order by absorbing loop
divergences in the renormalization of parameters of higher order
counterterms, known as low energy constants (LECs)
 that \emph{carry no energy or mass dependence}.
Their  values depend
on the specific QCD dynamics, and have to be determined either from
experiment or from QCD.
The relevant remark for us is that, up to 
the desired order,  the ChPT expansion
provides a {\it systematic and model independent} 
description of how meson masses and amplitudes depend on QCD
parameters like the light quark masses $\hat m=(m_u+m_d)/2$ and $m_s$,
or the leading $1/N_c$ behavior \cite{'tHooft:1973jz}.

% Despite ChPT is only valid at very low energies and,
% being an expansion, cannot generate resonance poles,
% it can be used to determine the subtraction constant of a disperion
% relation that describes amplitudes up to the resonance region. 

\vspace*{-.3cm}
\section{Dispersion relations and unitarization}
\vspace*{-.2cm}

Elastic resonances appear as poles on the
second Riemann sheet of the meson-meson
scattering partial waves $t_{IJ}$ of definite isospin $I$ 
and angular momentum $J$. 
At physical values of $s$, elastic unitarity implies 

\begin{equation}
  \ima t_{IJ}(s)=\sigma (s) \vert t_{IJ}(s)\vert^2 
  \;\;\Rightarrow\;\; 
  \ima\frac1{t_{IJ}(s)}=-\sigma(s), \quad t_{IJ}=\frac{1}{\rea t_{IJ}^{-1} - i \sigma},
\qquad {\rm with}
  \quad \sigma(s)=2p/\sqrt{s},
  \label{unit}
\end{equation}
where $s$ is the Mandelstam variable and $p$ is the
center of mass momentum. However, 
ChPT amplitudes, being an expansion
$t_{IJ}\simeq t_{IJ}^{(2)}+t_{IJ}^{(4)}+\cdots$, with
$t^{2k}=O(p^{2k})$, can only satisfy
Eq. (\ref{unit}) perturbatively
\begin{equation}
  \label{unitpertu}
  \ima t_{IJ}^{(2)}(s)=0,\;\;\;
  \ima t_{IJ}^{(4)}(s)=\sigma(s)\vert t_2(s)\vert^2,\;\;\dots
  \quad\Rightarrow\quad
  \ima t_{IJ}^{(4)}(s)/t_{IJ}^{(2)\,2}(s)=\sigma(s),
\end{equation}
and cannot generate poles. Despite the resonance region lies
beyond the reach of standard ChPT, it can be
reached combining ChPT with dispersion theory either for the
amplitude~\cite{gilberto}
or for the inverse amplitude through the
Inverse Amplitude Method (IAM)~\cite{Truong:1988zp,Dobado:1992ha,GomezNicola:2007qj}. The first approach
has been successfully used, combined with data on other channels and high energies, to,
for instance, determine precisely the parameters of the $f_0(600)$ or $\kappa(800)$ resonances.
Unfortunately, this additional experimental input makes it difficult to
relate these results to QCD parameters 
like $N_c$ or $\hat m$. 
Hence
we will concentrate on the {\it one-channel} IAM 
\cite{Truong:1988zp,Dobado:1992ha}, since it uses ChPT 
only up to a given order inside 
a dispersion relation -- without additional input or further model dependent assumptions -- providing an elastic
unitary amplitude with the correct ChPT 
expansion up to that order. Other unitarization techniques will be commented below.
\vspace*{-.5cm}

\vspace*{-.4cm}
\subsection{The one-loop ChPT Inverse Amplitude Method}
\vspace*{-.2cm}

For a partial wave $t_{IJ}(s)$, we can write a
dispersion relation (subtracted three times to suppress high energy contributions)
\begin{equation}
t_{IJ}(s)=C_0+C_1s+C_2s^2+
\frac{s^3}\pi\int_{s_{th}}^{\infty}\frac{\ima
t_{IJ}(s')ds'}{s'^3(s'-s-i\epsilon)} + LC(t_{IJ}).
\label{disp}
\end{equation}Note we have explicitly
written the integral over the right hand cut (or physical cut, 
extending from threshold, $s_{th}$ to infinity) 
but we have abbreviated by $LC$ the equivalent
expression for the left cut (from 0 to $-\infty$). We could do similarly with 
other cuts, if present, as in the $\pi K$ case.
Similar expressions hold for $t^{(2)}$ and $t^{(4)}$, 
but remembering that $t^{(2)}$ is a pure tree level amplitude and 
it does not have imaginary part nor cuts, they read:
\begin{eqnarray}
t_{IJ}^{(2)} = a_0+a_1s, \qquad  %\nonumber \\
t_{IJ}^{(4)} = b_0+b_1s+b_2s^2+     
\frac{s^3}\pi\int_{s_{th}}^{\infty}\frac{\ima
t_{IJ}^{(4)}(s')ds'}{s'^3(s'-s-i\epsilon)}+LC(t^{(4)}_{IJ}).
\label{disp1}
\end{eqnarray}
Note that from Eq.\eqref{unit}
the imaginary part of the {\it inverse amplitude} is {\it exactly} known in the elastic regime.
We can then write a dispersion relation like that in \eqref{disp} 
but now for the auxiliary function $G=(t_{IJ}^{(2)})^2/t_{IJ}$, i.e.,
\begin{equation}
G(s)=G_0+G_1s+G_2s^2+     \\   \nonumber
\frac{s^3}\pi\int_{s_{th}}^{\infty}
\frac{\ima G(s')ds'}{s'^3(s'-s-i\epsilon)}+LC(G)+PC,
\label{Gdisp}
\end{equation}
where now $PC$ stands for possible pole contributions in $G$ coming from 
zeros in $t_{IJ}$. It is now straightforward to expand the subtraction constants
and use that
 $\ima t_{IJ}^{(2)}=0$ and  $\ima t_{IJ}^{(4)}=\sigma\vert t_{IJ}^{(2)}\vert^2$,
so that $\ima G= -\ima t_{IJ}^{(4)}$. In addition,
 up to the given order, $LC(G)\simeq -LC(t_{IJ}^{(4)})$,
whereas $PC$ is of higher order and can be neglected on a first stage. Then
\begin{eqnarray}
\frac{t_{IJ}^{(2)2}}{t_{IJ}}\simeq a_0+a_1s-b_0-b_1s-b_2s^2 %\nonumber  \\   
-\frac{s^3}\pi\int_{s_{th}}^{\infty}\frac{\ima
t_{IJ}^{(4)}(s')ds'}{s'^3(s'-s-i\epsilon)}-LC(t_{IJ}^{(4)})
\simeq t_{IJ}^{(2)}-t_{IJ}^{(4)}.
\label{preIAM}
\end{eqnarray}
We have thus arrived to the so-called IAM:
\begin{equation}
t_{IJ}\simeq
t_{IJ}^{(2)2}/(
t_{IJ}^{(2)}-t_{IJ}^{(4)} ),
\label{IAM}
\end{equation}
that provides an elastic amplitude satisfying unitarity and has the correct
low energy expansion of ChPT up to the order we have used. 
The $PC$ contribution has been calculated  explicitly \cite{GomezNicola:2007qj} and 
shown to be, not just formally suppressed, but numerically negligible
except near the Adler zeros, away from the physical region.
It is straightforward
to extend the IAM to other elastic channels or higher orders \cite{Dobado:1992ha}.
Naively, by looking at \eqref{unit}, it seems that the IAM 
is derived by replacing $\rea t_{IJ}^{-1}$ by its $O(p^4)$ ChPT expansion.
But, strictly speaking,
 \eqref{unit} is only valid in the real axis, whereas our derivation allows
us to consider the amplitude in the complex plane, and, 
in particular, look for poles of the associated resonances.
Let us remark that ChPT has been used
always \emph{at low energies} to evaluate parts of a
dispersion relation, whose elastic unitarity cut is
taken into account exactly. Thus, the IAM formula is
reliable up to energies where inelasticities become
important (even though ChPT does not converge at those
energies) because ChPT is not being used there.
Only when the energy is close to the Adler zero one should use
a slightly modified version of the IAM \cite{GomezNicola:2007qj}.
When reexpanding, a few of the higher order terms are produced correctly
by the unitarization but not the complete series--- for
a discussion of this issue for the scalar pion form factor see 
Ref.\cite{Gasser:1990bv}.

\begin{figure}
%\begin{tabular}{c}
\vbox{
\centering
  \includegraphics[scale=.45,angle=-90]{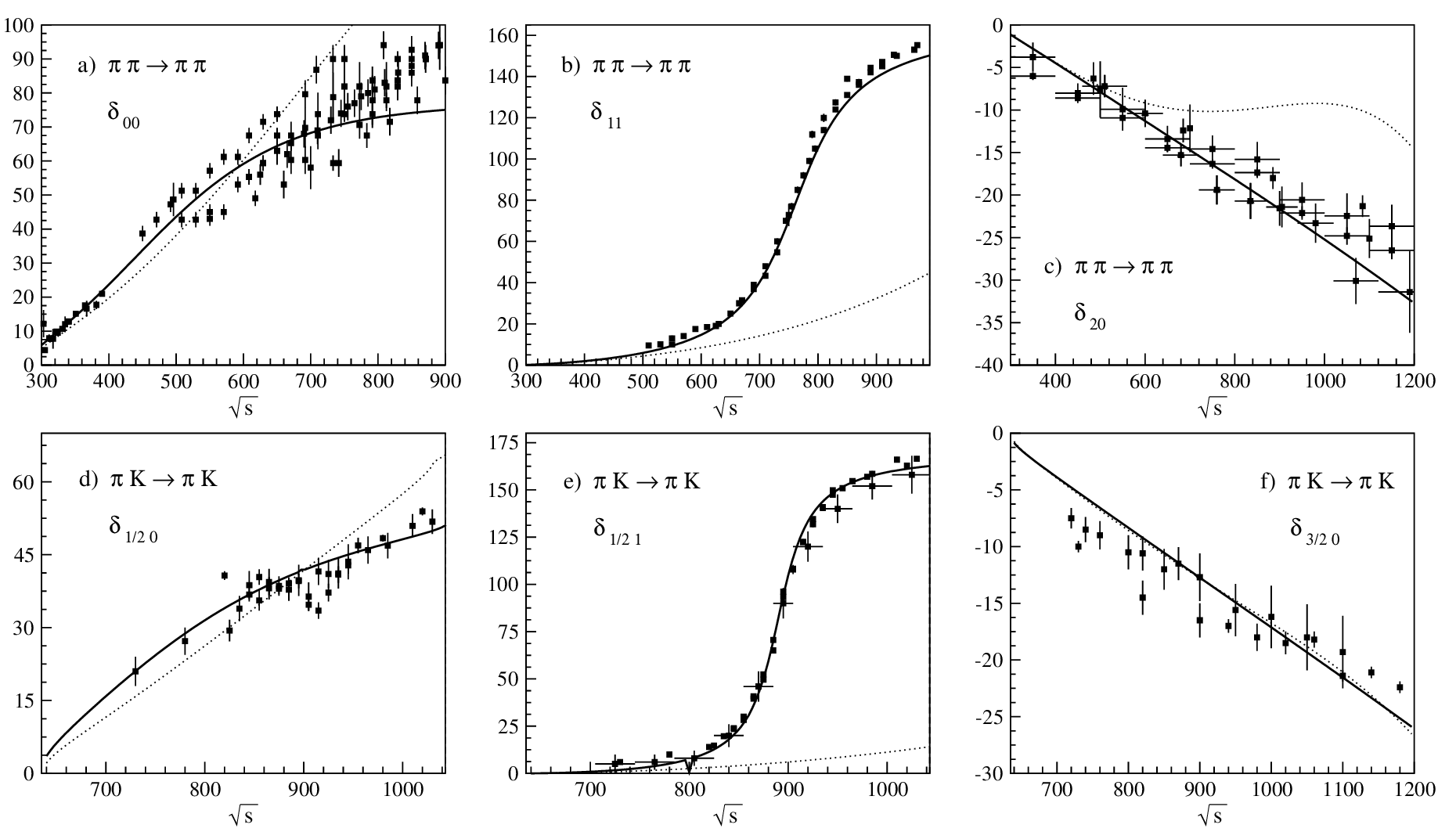}\\
  \includegraphics[scale=.45,angle=-90]{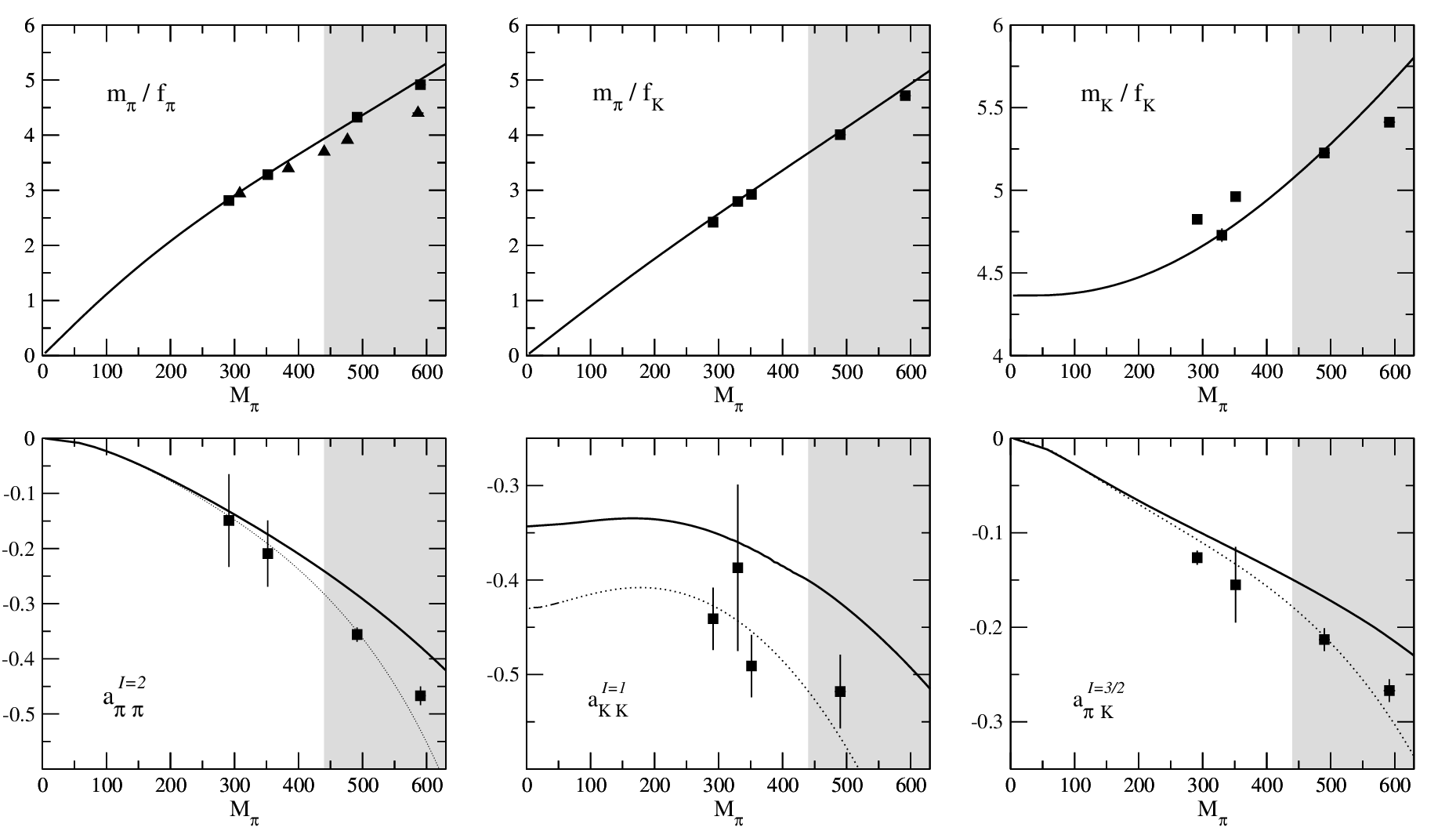}
}
%\end{tabular}
\caption{
  Results of our updated IAM fit \cite{Jeni} (continuous line). We also show
non-unitarized ChPT results with the LECs given in
the second reference of \cite{GomezNicola:2001as} (dot-dashed line). . 
Two upper rows: IAM versus experimental data on $\pi\pi$ and $\pi K$
scattering. Two lower rows: fit results 
compared to lattice calculations  \cite{lattice} of $f_\pi$, $f_K$,
$m_\pi/f_\pi$ and the $\pi^+\pi^+$, $K^+ K^+$, $K^+\pi^+$ 
scattering lengths. We fit up to $m_\pi=440 MeV$, but even beyond
lattice results are well described (grey areas). Experimental
references are detailed in \cite{GomezNicola:2001as}.
}
  \label{fig:fits}
\end{figure}

In Fig.\ref{fig:fits}, we present some preliminary 
results \cite{Jeni} of an updated fit of the IAM
$\pi\pi$ and $\pi K$ scattering amplitudes to data, but simultaneously fitting
the available lattice results on $m_\pi, m_K, f_\pi, f_K$ 
and some scattering lengths.
It is important to remark that the resulting 
LECs are in fairly good agreement
with standard determinations: no fine tuning is required. 
As usual the $f_0(600)$,
$\rho(770)$, $\kappa(800)$ and $K^*(892)$ appear 
as poles in the second Riemann sheet of their corresponding partial wave.
 Actually,
already ten years ago \cite{Dobado:1992ha}, 
with the elastic IAM we were able to 
generate poles for the $\rho(770)$, $K^*(892)$ and 
the controversial $\sigma$ (or $f_0(600)$), 
{\it without any modeling of the integrals but just ChPT approximations}.
The fact that resonances are 
\emph{not introduced by hand}, but generated
from first principles and data, is relevant because the 
existence and nature of scalar resonances is the
subject of a long-lasting intense debate. The fact that 
{\it the only input parameters are those of ChPT} is very relevant because we then know
how to relate our amplitudes to QCD parameters like $N_c$ or the quark masses.
\vspace*{-.3cm}

\subsection{Other unitarization techniques within the coupled channel formalism}
\vspace*{-.3cm}

Naively one can arrive to \eqref{IAM} in a matrix form, ensuring
coupled channel unitarity, just by expanding
the real part of the inverse $T$ matrix. 
Unfortunately, {\it there is still no dispersive derivation}
including a left cut {\it for the coupled channel case}.
Being much more complicated,  different approximations to $\rea T^ {-1}$
have been used:

$\bullet$ The fully renormalized one-loop ChPT calculation of $\rea T^ {-1}$ 
provides the correct ChPT expansion in all channels,
also with left cuts approximated to $O(p^4)$ \cite{Guerrero:1998ei,GomezNicola:2001as}. 
Indeed, using
LECs consistent with previous determinations within standard ChPT,
it was possible \cite{GomezNicola:2001as}
to describe below 1.2 GeV all the scattering
channels of two body states made of pions, kaons or etas. Simultaneously,
this approach \cite{GomezNicola:2001as} generates poles associated to the $\rho(770)$
and $K^*(892)$ vector mesons,
together with the  $f_0(980)$, $a_0(980)$, $f_0(600)$ 
and $\kappa$ (or $K_0(800)$) scalar resonances.

$\bullet$ Originally \cite{Oller:1997ng}, the coupled channel
IAM  was used neglecting the crossed loops and tadpoles.
This approach is considerably 
simpler, and although
 it is true that the left cut is absent, its numerical influence
was shown to be rather small, since the meson-meson data
are nicely described with very reasonable
chiral parameters and generates all the poles enumerated above. 
Let us remark that this approximation keeps the s-channel loops
but also the tree level up to $O(p^4)$, and that this tree level
encodes the effect of  heavier resonances, like the rho. Thus,
contrary to some common belief, this approach still incorporates,
for instance,
the low energy effects of t-channel rho exchange.

$\bullet$ Finally, if only scalar meson-meson scattering is of interest, it is possible to 
use just one cutoff (or a dimensional regularization scale or a subtraction constant)
that numerically mimics 
the combination of chiral parameters that appear in those 
scalar channels. This method -- known as the "`chiral unitary approach"'-- has become very popular, 
even beyond the meson-meson interaction realm,
due to its great simplicity but remarkable success
\cite{Oller:1997ti} and also because it is rather simple to
relate to the Bethe-Salpeter formalism \cite{Nieves:1998hp} that provides additional physical insight on unitarization.

With this method it has been shown \cite{Oller:2003vf} that, in the SU(3) limit, and assuming no quark mass dependence on the cutoff, all light scalar resonances degenerate into an octet and a singlet.

Also with this method, but using a chiral
 Lagrangian for the pseudoscalar-vector interaction
it has been possible to generate axial-vector mesons \cite{Lutz:2003fm}.
\vspace*{-.2cm}

\section{The nature of resonances from their leading $1/N_c$ behavior}
\vspace*{-.2cm}

The QCD $1/N_c$ expansion \cite{'tHooft:1973jz}, valid
 in the whole energy region, 
provides a rigorous definition of $\bar qq$ bound states: their 
masses and widths behave as $O(1)$ and $O(1/N_c)$, respectively.
The QCD leading $1/N_c$ behavior of $f_\pi$
 and the LECs is well known, and ChPT amplitudes 
have no cutoffs or subtraction constants
where spurious $N_c$ dependences could hide.
Hence, by scaling with $N_c$ the ChPT parameters in the IAM, 
the mass and width $N_c$ dependence of the resonances has been determined
to one and two loops
 \cite{Pelaez:2003dy,Pelaez:2006nj}.
These are
defined from the pole position as $\sqrt{s_{pole}}=M-i\Gamma$.
However, {\it a priori}, one should be careful {\it not to take $N_c$ too large, and in particular
to avoid the $N_c\to\infty$ limit, because it is a weakly interacting limit}. As shown above, the IAM relies
on the fact that the exact elastic $RC$ contribution dominates the
dispersion relation.
Since the IAM describes data and the resonances within, say, 10 to 20\% errors, this means that at $N_c=3$ the other contributions are not
approximated badly.  But meson loops, responsible for the $RC$, scale as $3/N_c$ whereas the inaccuracies due to the approximations scale partly as $O(1)$.
Thus, we can estimate that those 10 to 20\% errors at $N_c=3$ may become 100\% errors at, say $N_c\sim30$ or $N_c\sim15$, respectively.
Hence we have never shown results \cite{Pelaez:2003dy,Pelaez:2006nj} beyond $N_c=30$, and even beyond $N_c\sim15$ they should be interpreted with care.
Of course, in special cases the IAM could still work for very large $N_c$,
as it is has been shown for the vector channel \cite{Nieves:2009ez}. 
But that is not the case for the scalar channel, which, 
if used for too large $N_c$, leads to inconsistencies \cite{Nieves:2009ez}
for some values of the LECs.

\begin{figure}[t]
  \centering
  \vbox{
    \hbox{
      \includegraphics[scale=.43]{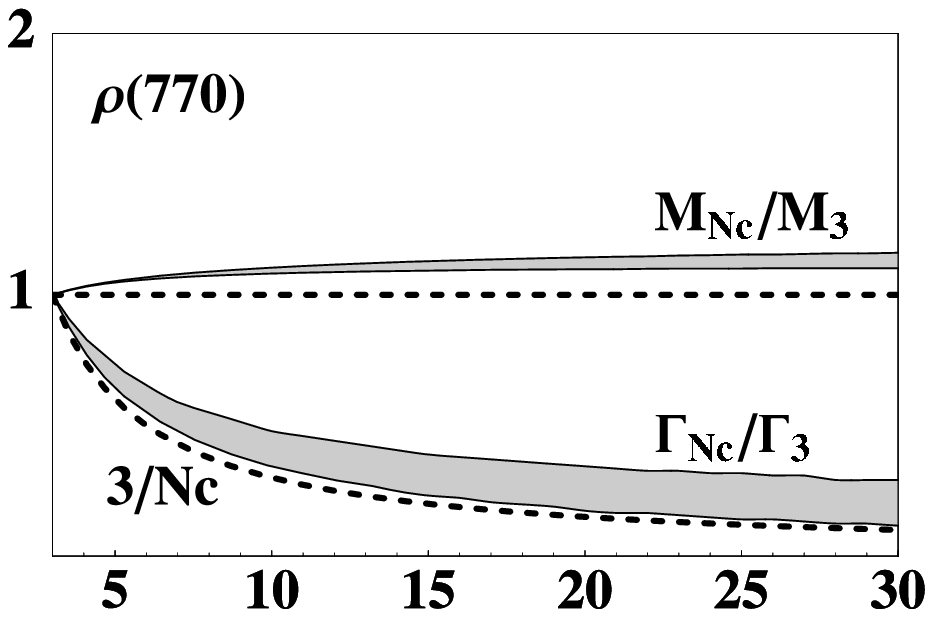}
      \includegraphics[scale=.43]{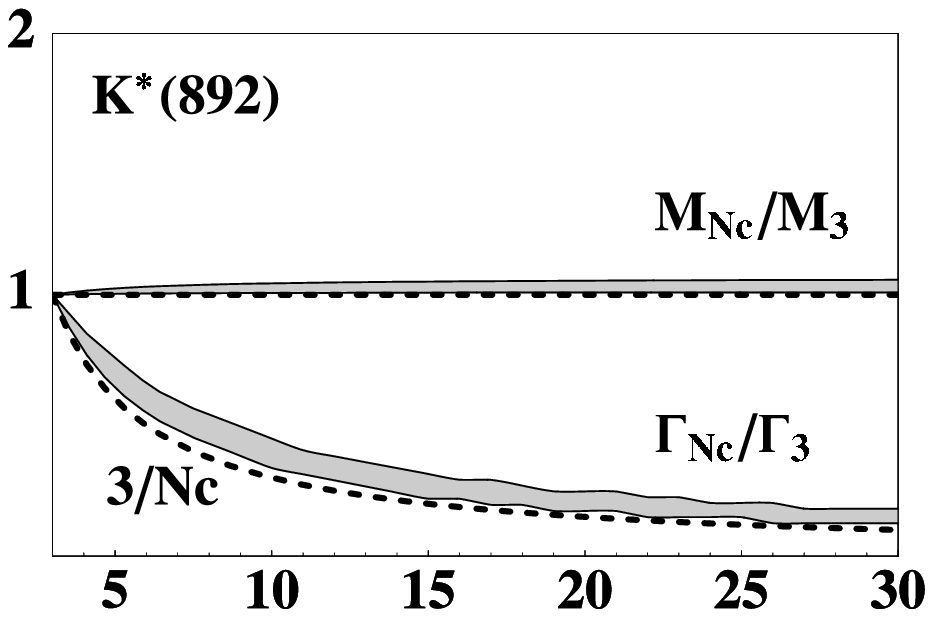}
      \includegraphics[scale=.43]{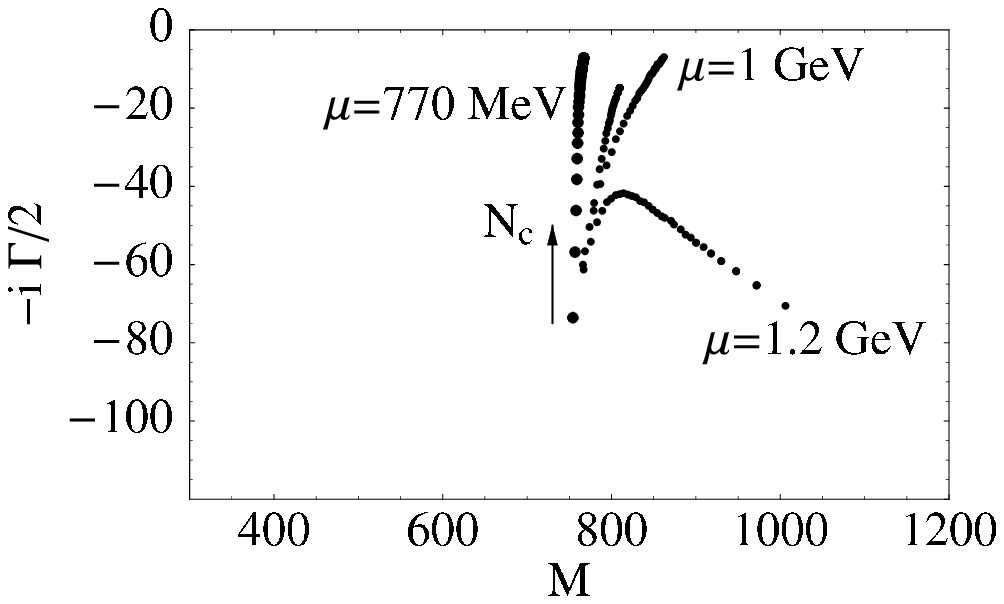}
    }
    \hbox{
      \includegraphics[scale=.43]{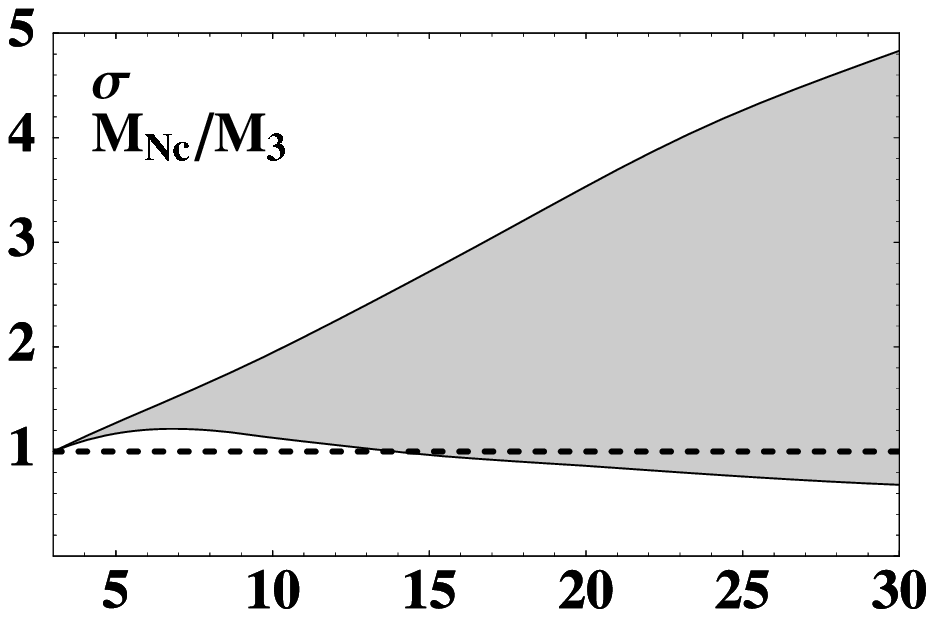}
      \includegraphics[scale=.43]{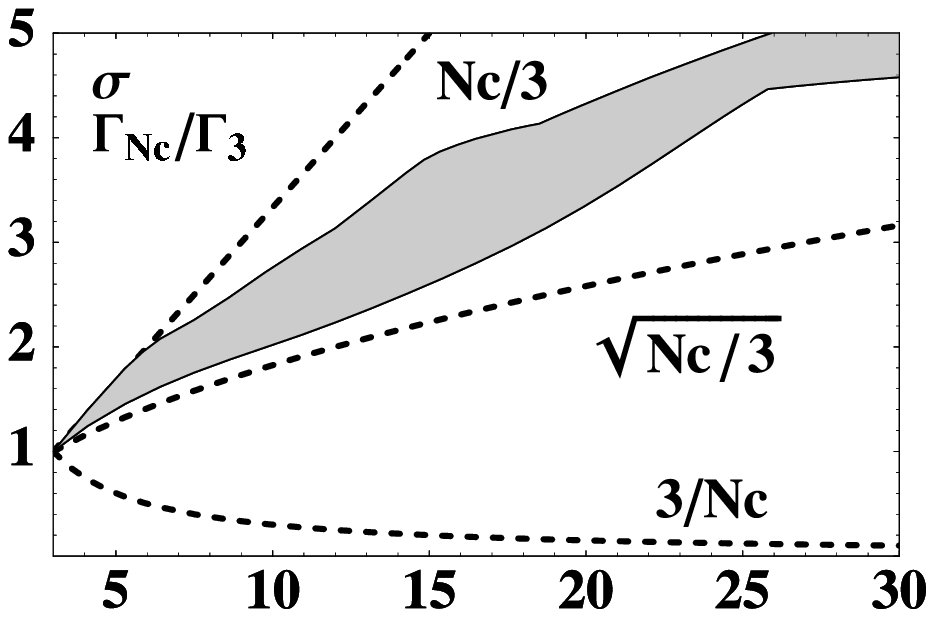}
      \includegraphics[scale=.43]{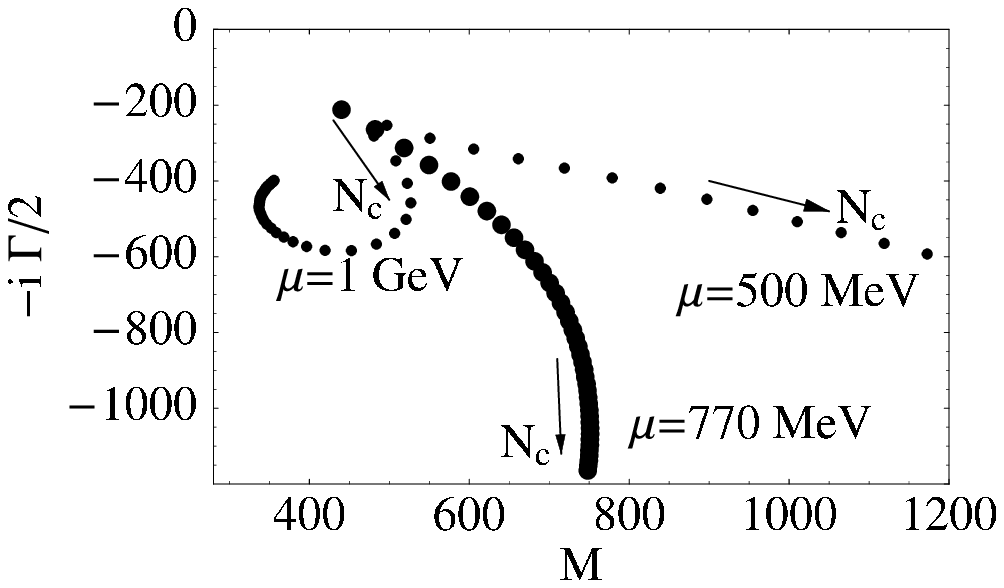}
    }
  }
  \caption{{\bf Top:} (Left and center) $N_c$ behavior of the $\rho$ and $K^*$
  mass and width. (Right) Different $\rho$ pole
  trajectories for different values of $\mu$, note that for
  $\mu=1.2$ GeV the $\rho$ pole goes away the real axis.
  {\bf Bottom:} (Left and center) $N_c$ behavior of the $\sigma$ mass and width. 
(Right) Different $\sigma$ 
  pole trajectories for different $\mu$ values.}
  \label{ncSU3}
\end{figure}

Thus, Fig.\ref{ncSU3} shows the
behavior of the $\rho$, $K^*$ and $\sigma$ masses and widths
found in  \cite{Pelaez:2003dy}. The $\rho$ and
$K^*$ neatly follow the expected behavior for 
a $\bar qq$ state: $M\sim 1$, $\Gamma\sim 1/N_c$.
The bands cover the uncertainty
in $\mu\sim 0.5-1$ GeV
 where the LECs are scaled with $N_c$.
Note also in Fig.2~(top-right) that, for that set of LECs, {\it outside this $\mu$ range}
the $\rho$ meson starts deviating from a 
a $\bar qq$ behavior. Something similar occurs to the $K^*(892)$.
Consequently, we cannot
apply the $N_c$ scaling at an arbitrary $\mu$ value,
if the well established $\rho$ and $K^*$ $\bar qq$ nature is to be reproduced.

% \begin{figure}
%   \includegraphics[height=.3\textheight]{rhonc}
%   \caption{Picture to fixed height}
% \end{figure}

In contrast, the $\sigma$ shows a 
different behavior from that of a pure $\bar qq$:
\emph{near $N_c$=3} both its mass and width
grow with $N_c$, i.e. its pole moves 
away from the real axis. 
Of course,  far from $N_c=3$, and
for some choices of LECs and $\mu$,
the sigma pole might turn back to the real axis \cite{Pelaez:2006nj,Nieves:2009ez,Sun:2004de}, 
as seen in Fig.2 (bottom-right). 
But, as commented above, the IAM is less reliable for large $N_c$,
and at most this behavior only suggests that
there {\it might be} a subdominant $\bar q q$ component \cite{Pelaez:2006nj}.
In addition, we have to ensure that the LECs used fit data
and reproduce the vector $\bar qq$ behavior.

Since loops are important in determining the scalar pole position,
but are $1/N_c$ suppressed compared to tree level terms with LECs,
it is relevant to check the  $O(p^4)$ results
with an $O(p^6)$ IAM calculation. This was done
 within $SU(2)$ ChPT in \cite{Pelaez:2006nj}. 
We defined a $\chi^2$-like function to measure how close 
 a resonance is from a $\bar qq$ $N_c$  behavior.
First, we used that $\chi^2$-like function at $O(p^4)$ to show
 that it is not possible for the $\sigma$ to behave predominantly as a $\bar q q$
while describing simultaneously the data
and the $\rho$ $\bar qq$ behavior, thus
{\it confirming the robustness of the conclusions for $N_c$ close to 3}. 
Next, we obtained a $O(p^6)$ data fit -- where
the $\rho$ $\bar qq$ behavior was imposed -- whose $N_c$ behavior for the
$\rho$ and $\sigma$ mass and width is shown in Fig.3~(left and center). Note that
both $M_\sigma$ and $\Gamma_\sigma$ grow with $N_c$ near $N_c=3$,
confirming the $O(p^4)$ result of a non $\bar qq$ dominant component.
However, as $N_c$ grows further, between $N_c\sim8$ and $N_c\sim 15$, where we still
trust the IAM results, $M_\sigma$
becomes constant and $\Gamma_\sigma$ starts decreasing. 
This may hint to a \emph{subdominant $\bar qq$ component},
arising as loop diagrams become suppressed when $N_c$ grows.
Finally, and despite this scenario is disfavored since the $\rho$ starts deviating from its $\bar qq$ behavior, we checked how big this $\sigma$ $\bar qq$ component
can be made. Thus we forced the $\sigma$ to behave 
as a $\bar qq$ using the above mentioned $\chi^2$-like measure.
We found that in the best case -- Fig.3. (right) -- this subdominant $\bar qq$
component could become dominant around $N_c>6-8$, at best, but
always with an $N_c\to\infty$ mass  above roughly 1 GeV instead of its physical $\sim 450$ MeV value.

\begin{figure}[t]
  \centering
  \hbox{
    \includegraphics[angle=-90,scale=.34]{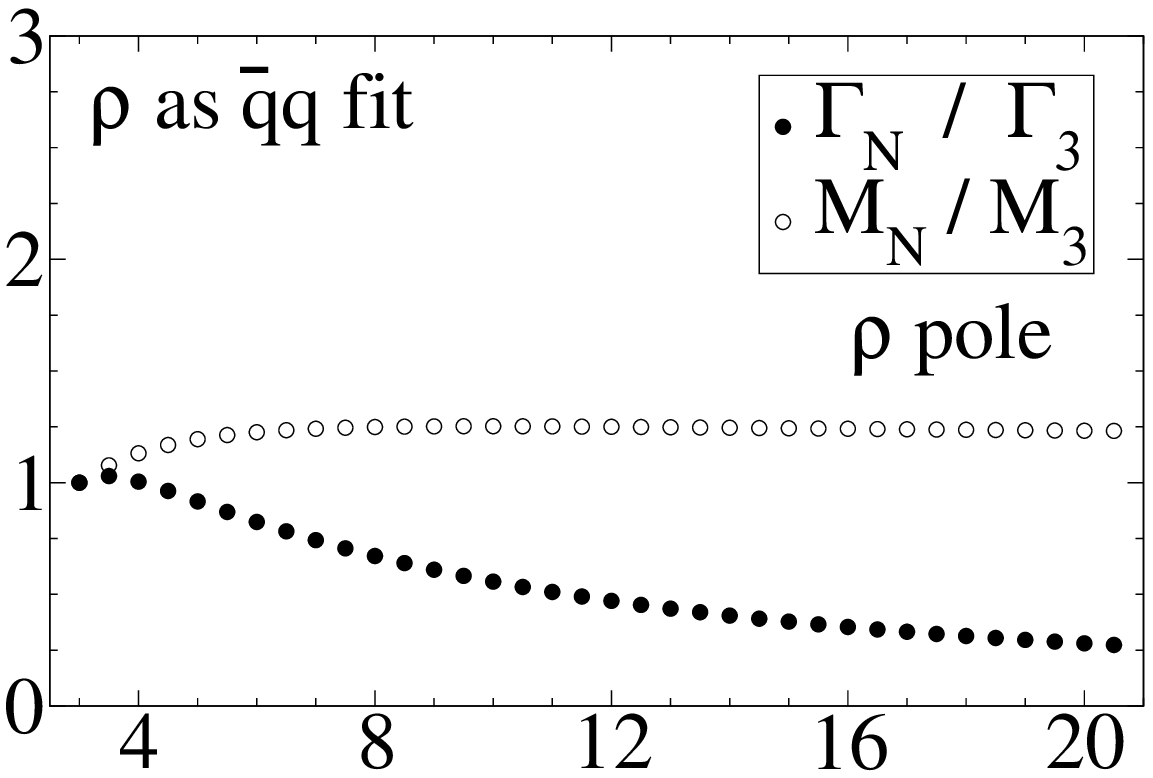}
    \includegraphics[angle=-90,scale=.34]{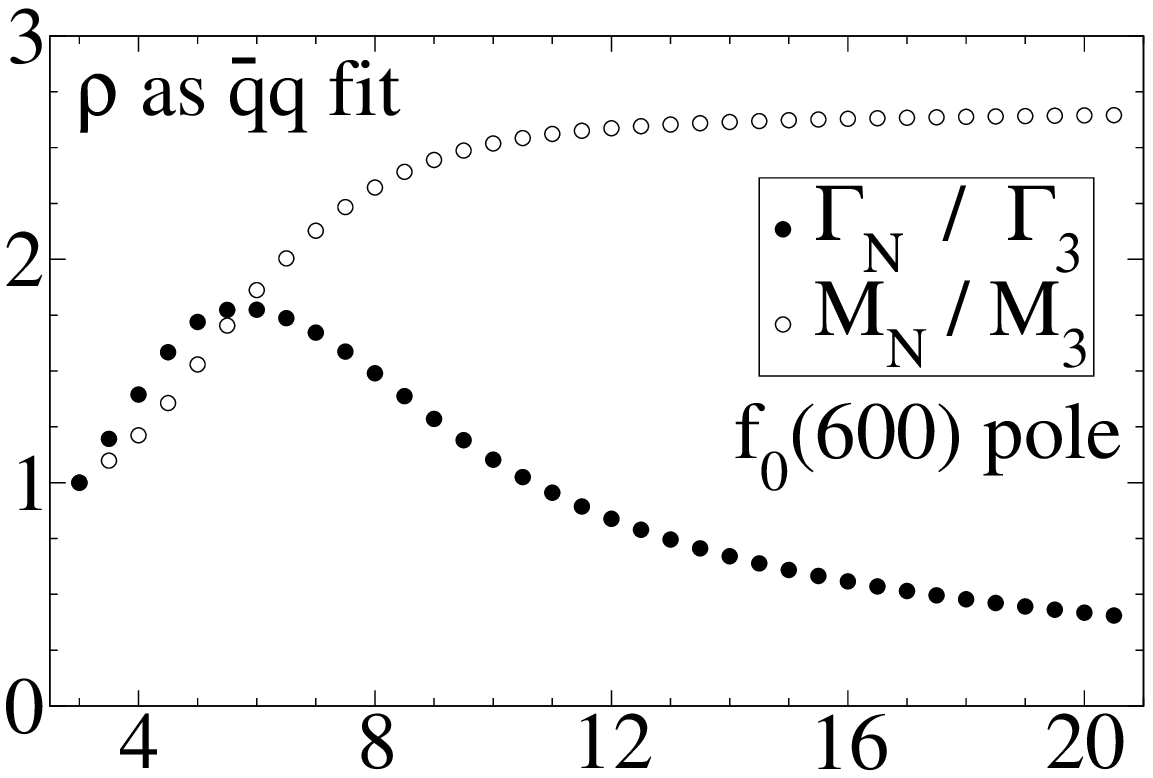}
    \includegraphics[angle=-90,scale=.34]{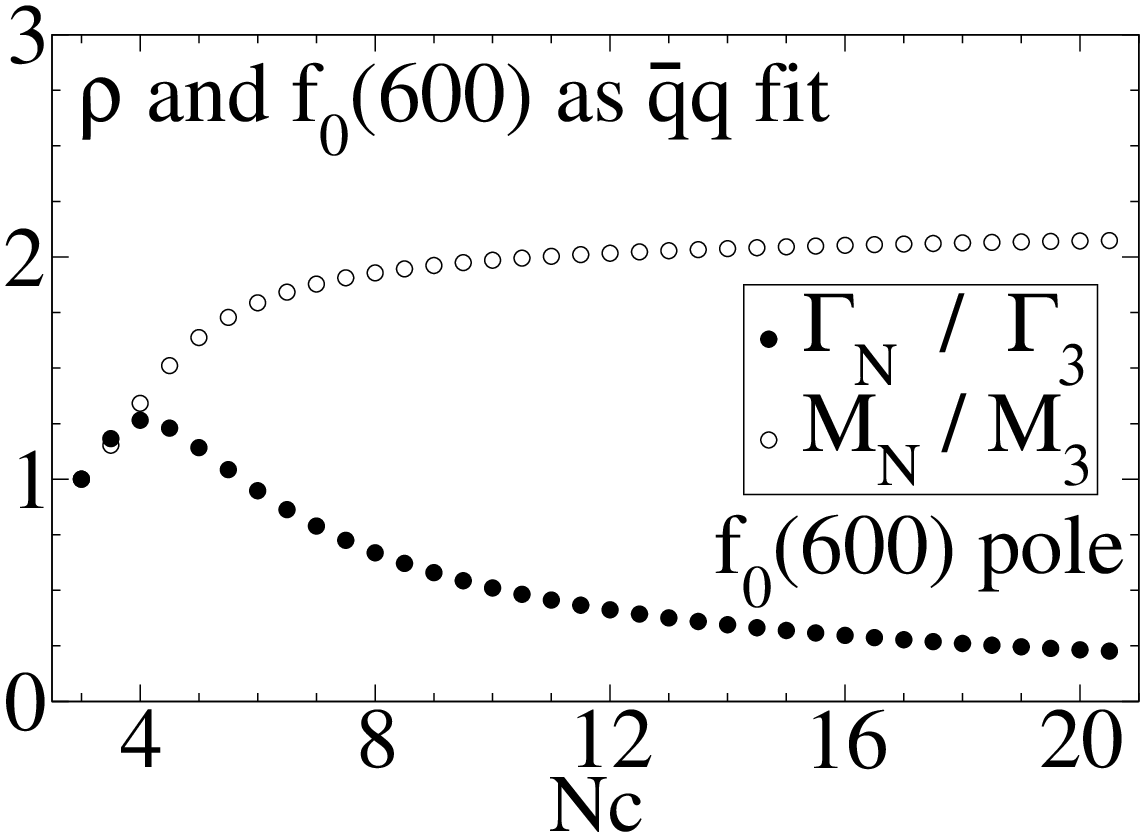}
  }
  \caption{{\bf Left and center}: $N_c$ behavior of the $\rho$
    and $\sigma$ pole at $O(p^6)$ with the ``$\rho$ as 
    $\bar qq$ fit''.
    {\bf Right:} Sigma behavior with $N_c$ at $O(p^6)$ with
    the ``$\rho$ and $\sigma$ as $\bar qq$ fit''.}
  \label{2loops}
\end{figure}

Let us emphasize again \cite{Pelaez:2005fd}
what can and {\it what cannot} be concluded from our results
and clarify some frequent questions and doubts raised in this and other meetings,
private discussions and the literature:
\newcounter{input}
\begin{list}{$\bullet$}{
\setlength{\leftmargin}{0.2cm}
\setlength{\labelsep}{0.cm}}

\vspace{-.2cm}
\item {\it The \underline{dominant component}  of the $\sigma$ and $\kappa$ {\it in meson-meson scattering} does not behave as a $\bar{q}q$}. Why ``dominant''?
    Because, most likely, scalars are a mixture of different
    states.  If the $\bar{q}q$ was {\it dominant}, they would behave
    as the $\rho$ or the $K^*$ in Fig.2.  {\it But a smaller fraction of $\bar{q}q$ cannot be
      excluded} and is somewhat favored in our $O(p^6)$ analysis \cite{Pelaez:2006nj}.
      
%\vspace{-.2cm}
  \item {\it Two meson and some tetraquark states \cite{Jaffe} have a consistent
      ``qualitative'' behavior}, i.e., both disappear in
    the  meson-meson scattering continuum as $N_c$
    increases. Our results are not able yet
     to establish the nature of that dominant component.
To do so other tools might be necessary
                         as, for instance, those outlined in \cite{Weinberg,baru}.
The most we could state is that the behavior of
      two-meson states or some tetraquarks might be
qualitatively consistent.
\end{list}

The $N_c\rightarrow\infty$
limit has been studied in \cite{Sun:2004de,Nieves:2009ez}. Apart from its
mathematical interest, it could have some physical
relevance if the data and the large $N_c$ uncertainty on the
choice of scale were more accurate. Nevertheless:
\begin{list}{\roman{input}.}{\usecounter{input}
\setlength{\leftmargin}{0.2cm}
\setlength{\labelsep}{0.cm}}
\vspace{-.2cm}
\item[$\bullet$ ] As commented above, {\it a priori the IAM is not reliable in the  $N_c\rightarrow\infty$ limit}, since it corresponds to a weakly interacting theory, where 
exact unitarity
becomes less relevant in confront of other approximations made in the IAM derivation. It has been shown \cite{Nieves:2009ez} that it might work
well in that limit in the vector channel of QCD but not in the scalar channel.

%\vspace{-.2cm}
\item[$\bullet$ ] Another reason to limit ourselves to $N_c$ not too
far from 3 is that in our calculations we have not included the $\eta'(980)$,
whose mass is related to the $U_A(1)$ anomaly and scales as 
$\sqrt{3/N_c}$.
Nevertheless, if in our calculations we keep $N_c<30$, its mass
would be $>310\,$MeV and thus pions are still the only relevant degrees
of freedom for the scalar channel in the $\sigma$ region.

%\vspace{-.2cm}
\item[$\bullet$ ] 
{\it Contrary to the leading $1/N_c$ behavior 
\underline{in the vicinity of  $N_c=3$},
the $N_c\rightarrow\infty$ 
limit does not give information on the ``dominant component''
of light scalars.} The reason was 
commented above: In contrast to $\bar{q}q$
states, that become bound, 
 two-meson and some tetraquark
states dissolve in the 
continuum as $N_c\rightarrow\infty$. 
Thus, even if we started with an infinitesimal $\bar{q}q$ component
in a resonance, for a sufficiently large $N_c$ 
it may become dominant, and beyond 
that $N_c$ the associated pole
would behave as a $\bar{q}q$ state.
Also, since the mixings of 
different components could change
with $N_c$, a too large $N_c$ could alter significantly
the original mixings. 
\end{list}

Actually, this is what happens for the one-loop IAM $\sigma$ resonance 
for $N_c\to\infty$, but 
it does {\it not} necessarily mean that 
the ``correct interpretation [...] is that
the $\sigma$ pole is a conventional 
$\bar{q}q$ meson environed by heavy pion clouds'' \cite{Sun:2004de}.
That the scalars are not conventional, is simply seen by comparing
them in Figs.1 and 2 with the ``conventional'' $\rho$ and $K^*$ in
those very same figures.
A large two-meson component is consistent,
but the $N_c\rightarrow\infty$ of the one-loop unitarized ChPT 
pole in the scalar channel
 limit is not unique \cite{Sun:2004de,Nieves:2009ez} 
given the uncertainty
in the chiral parameters. Moreover, for some LECs the scalar channel one-loop
IAM in the $N_c\rightarrow\infty$ limit can lead to phenomenological inconsistencies 
\cite{Nieves:2009ez}, since poles can even move to negative 
 squared mass values (weird), 
to infinity or to a  positive mass square. 
That is one of the reasons why in the figures here and in \cite{Pelaez:2003dy,Pelaez:2006nj}
we only plot up to 
$N_c=30$, but not 100, or a million.
Hence, robust 
conclusions on the dominant light scalar component
can be obtained not too far
 from real life, say $N_c<15$ or 30,  for a $\mu$ choice between
roughly $0.5$ and 1 GeV, that simultaneously ensures the $\bar qq$ dependence for the $\rho$ and $K^*$ mesons.  Note, however, that under these same 
conditions the two-loop IAM still finds, not only 
a dominant non-$\bar qq$ component, but also a hint of a $\bar qq$ subdominant component, which
 is not conventional in the sense that it appears at a much higher mass than the physical $\sigma$.
This may support the existence of a second $\bar qq$ scalar octet above 1 GeV \cite{VanBeveren:1986ea}. 

Finally, using not the IAM, but the chiral unitary approach with a natural 
range for the cutoff $N_c$ dependence, it has also been suggested \cite{Geng:2008ag} that a 
large, in some cases dominant, non $\bar qq$ behavior could exist in axial vector mesons.
\vspace*{-.2cm}

\section{Quark mass dependence of resonances}
\vspace*{-.2cm}

ChPT provides a rigorous expansion of meson masses in terms of quark masses
(at leading order $M_{meson}^2\sim m_q$). Thus, 
by changing the meson masses in the amplitudes, we see how the poles
generated with the IAM depend on quark masses.
In \cite{Hanhart:2008mx} we presented the SU(2) analysis for the $\rho$ and $\sigma$
but here we also report on our recent developments \cite{Jeni} in the SU(3) formalism and
the $\kappa(800)$ and $K^*(892)$ strange resonances.

The values of $m_\pi$ considered should fall within the ChPT
range of applicability and allow for some elastic $\pi\pi$ and $\pi K$
regime below $K\bar K$ or $K\eta$ thresholds, respectively. Both criteria are
satisfied if $m_\pi\leq 440$ MeV, since $SU(3)$ ChPT
still works with such kaon masses, and because for
$m_\pi\simeq 440$ MeV, the kaon mass becomes $\simeq 600$ MeV. 
Of course, we expect higher order
corrections, which are not considered here, to become more
relevant as $m_\pi$ is increased. Thus, our results become less
reliable as $m_\pi$ increases due to the $O(p^6)$ 
corrections which we have neglected

\begin{figure}[t]
   \centering
     \hbox{
       \includegraphics[scale=0.3,angle=0]{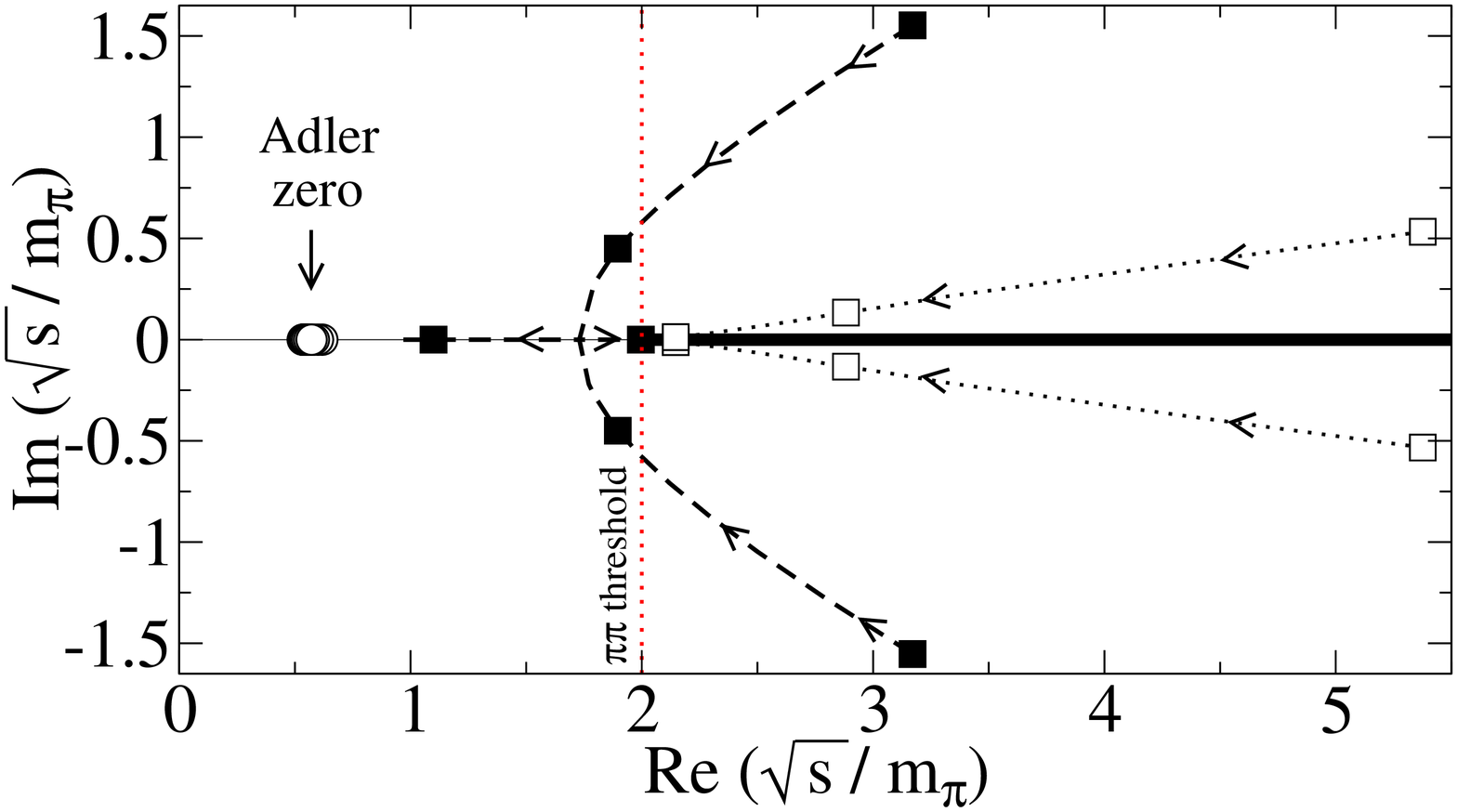}
       \includegraphics[scale=0.68,angle=0]{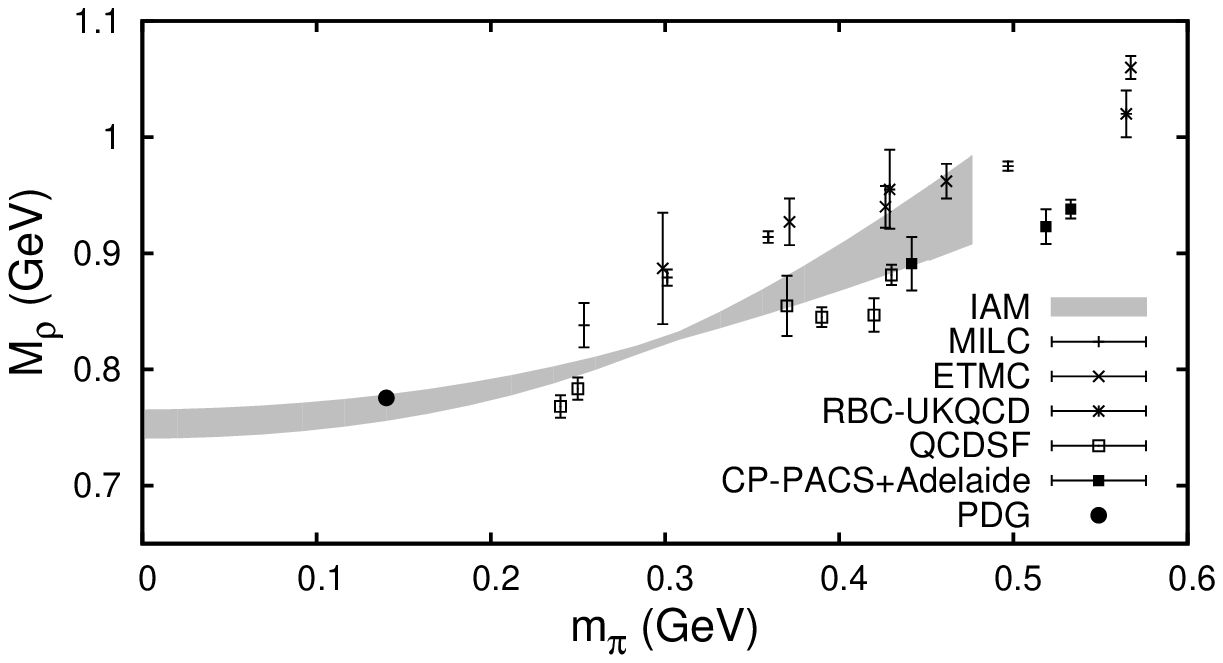}
     }
     \caption{ {\bf Left:} Movement of the $\sigma$ (dashed lines) 
       and $\rho$ (dotted
       lines) poles for increasing $m_\pi$ (direction indicated by the
       arrows) on the second sheet.  The filled (open) boxes denote the
       pole positions for the $\sigma$ ($\rho$) at pion masses $m_\pi=1,\
       2,$ and $3 \times m_\pi^{\rm phys}$, respectively. For
       $m_\pi=3m_\pi^{\rm phys}$ three poles accumulate in the plot 
       very near the $\pi\pi$ threshold. Note that all poles are
       always far enough from the Adler zero (circles).
       {\bf Right:} Comparison of our results for the $M_\rho$
       dependence on $m_\pi$ with some recent lattice results
       from \cite{lattice1}. The grey band covers only the
       error coming from the LECs uncertainties.}
       \label{poles}
%   \end{center}
 \end{figure}

Fig. \ref{poles} (left) shows the evolution of the $\sigma$ and $\rho$
pole positions as $m_\pi$ is increased. In order to see the pole
movements relative to the two pion threshold, which is also increasing,
all quantities are given in units of $m_\pi$, so the threshold is
fixed at $\sqrt{s}=2$. Both poles move closer to threshold and
they approach the real axis. The $\rho$ poles reach the real axis
as the same time that they cross threshold.
One of them jumps into the first sheet and stays below
threshold in the real axis as a bound state, while its conjugate
partner remains on the second sheet practically at the very same
position as the one in the first. In contrast, the $\sigma$
poles go below threshold with a finite imaginary part before they
meet in the real axis, still on the second sheet, becoming
virtual states. As $m_\pi$ is increased further, one of the poles
moves toward threshold and jumps through the branch point to the
first sheet staying in the real axis below threshold, very
close to it as $m_\pi$ keeps growing. The other $\sigma$ pole moves
down in energies further from threshold and remains 
on the second sheet.
These  very asymmetric poles could be
a signal of a prominent molecular component \cite{Weinberg,baru},
at least for large pion masses.
Similar movements have been found within quark models
\cite{vanBeveren:2002gy} and a finite density analysis 
\cite{FernandezFraile:2007fv}.

Fig. \ref{poles} (right) shows our results for the $\rho$ mass
dependence on $m_\pi$ compared with some recent lattice 
results~\cite{lattice1}, and the
PDG value for the $\rho$ mass. 
Now the mass is defined as the point where the phase shift crosses $\pi/2$,
except for those $m_\pi$ values where the $\rho$ becomes a bound
state, where it is defined again from the pole position.
Taking into
account the incompatibilities within errors between
different lattice collaborations, we find a qualitative good
agreement with the lattice results. Also,
we have to consider that the $m_\pi$ dependence
in our approach is correct only up to NLO in ChPT, and
we expect higher order corrections to be important for
large pion masses. The $M_\rho$ dependence on $m_\pi$
 agrees nicely with the estimations 
for the two first coefficients of its chiral expansion \cite{bruns}.

\begin{figure}[t]
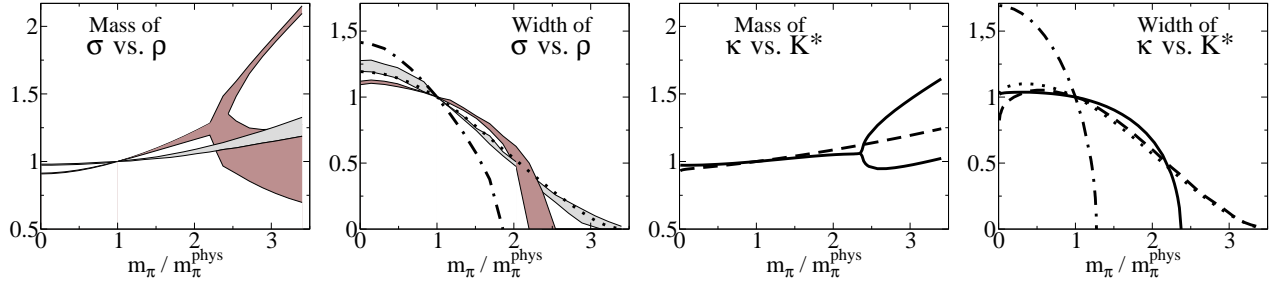

\centering
\begin{tabular}{c@{\hspace{0.2cm}}c@{\hspace{0.2cm}}c@{\hspace{0.2cm}}c}
  \includegraphics[scale=.39]{masasigmarho.eps}&
  \includegraphics[scale=.39]{anchosigmarho.eps}&
  \includegraphics[scale=.39]{masakappak.eps}&
  \includegraphics[scale=.39]{anchokappak.eps}
  \end{tabular}
     \caption{\label{massandwidth} $m_\pi$ dependence of resonance masses
       and widths in units of the physical values. 
       In the two left panels the dark (light) band shows the results for
       the $\sigma$ ($\rho$). The width of the bands reflects 
       the uncertainties
       induced only from the uncertainties in the SU(2) LECs.  
Similarly, in the two right panels we show our recent
developments \cite{Jeni} for the $\kappa(800)$ and $K^*(892)$
using central values
of the SU(3) IAM fits.
The dot-dashed (dotted)
       line shows the $m_\pi$ dependence of the corresponding scalar (vector) 
       width from the change of phase space only, assuming
       a constant coupling of the resonance to two mesons.
}
 \end{figure}

In Fig. \ref{massandwidth} (left) we compare the $m_\pi$ dependence
of $M_\rho$ and $M_\sigma$ (defined from the pole position
$\sqrt{s_{pole}}=M-i\Gamma /2$), normalized to their physical values.
The bands cover the LECs uncertainties. We see that both masses
grow with $m_\pi$, but $M_\sigma$ grows faster than $M_\rho$. 
Below $m_\pi\simeq 330$  MeV we only show one line because the two
conjugate $\sigma$ poles have the same mass. Above 330 MeV, these
two poles lie on the real axis with two different masses. The
heavier pole goes towards threshold and around $m_\pi\simeq 465$ MeV
moves into the
first sheet, but that is beyond our applicability limit.

In the next panel of Fig. \ref{massandwidth} we compare the $m_\pi$
dependence of $\Gamma_\rho$ and $\Gamma_\sigma$ normalized to their
physical values: note that both widths become smaller. 
We compare this decrease with the expected phase space reduction 
 as resonances approach the $\pi\pi$ threshold.
We find that $\Gamma_\rho$ follows very well
this expected behavior, which
implies that the $\rho\pi\pi$ coupling is almost $m_\pi$ independent.
In contrast, $\Gamma_\sigma$ deviates from the
phase space reduction expectation. This suggests a strong $m_\pi$
dependence of the $\sigma$ coupling to two pions, necessarily
present for molecular states \cite{baru,mol}.

Finally, in the last two panels of Fig.\ref{massandwidth}
we compare the mass and width dependence on $\hat m$
of the $\kappa(800)$ versus
the $K^*(892)$, keeping $m_s$ fixed. Note that the same pattern of the $\sigma-\rho$ system
is repeated. Belonging to the same octet, the $K^*(892)$ and $\rho$ behave very similarly, and 
both their widths follow just phase space reduction. 
The  $\sigma$  $\kappa$ behavior are only qualitatively similar,
the latter being somewhat softer. Among other effects,
this might be due to a possible significant admixture of singlet state in the $\sigma$.

\vspace*{-.3cm}

\section{Summary}
\vspace*{-.2cm}

We have reviewed how the Inverse Amplitude Method (IAM)
\cite{GomezNicola:2007qj} is derived from the first principles of
analyticity, unitarity, and Chiral Perturbation Theory (ChPT)  at low energies. 
It is able to generate, as
poles in the amplitudes,
the light resonances appearing in meson-meson elastic scattering, 
 without any a priori assumptions. Up to a given order in ChPT, it
yields the correct dependences on the quark masses and the number of colors. 

The resonance leading $1/N_c$ behavior suggests 
that the dominant component of light scalars
does not behave as a
$\bar{q}q$ state as $N_c$ increases not far from $N_c=3$.
When using the two loop IAM
result in SU(2), below $N_c\sim\,$15 or 30,  
there is a hint of a subdominant $\bar{q}q$ component, but
 arising at roughly twice the mass of the physical $\sigma$.

We have also predicted the evolution of the $f_0(600), \rho(770), \kappa(800)$ and $K^*(892)$
pole positions with
increasing pion (quark) mass \cite{Hanhart:2008mx,Jeni} and have seen how
they become bound states: softly in the vector case and with a non-analyticity in the scalar case. 
We have also shown that the
vector-meson-meson coupling constant is almost $m_\pi$ independent and we
have found a qualitative agreement with some lattice results for the
$\rho$ mass evolution with $m_\pi$.  These findings might be
relevant for studies of the meson spectrum and form factors---see
Ref.~\cite{Guo:2008nc} ---on the lattice. Work is in progress \cite{Jeni} to study also the strange quark mass dependence. 
\vspace*{-.5cm}

\section*{Acknowledgments}
\vspace*{-.2cm}

J.R.P. thanks the organizers of Hadron2009 for the invitation and
for their work to create such a pleasant but exciting conference.
Work partially supported by Spanish Ministerio de
Educaci\'on y Ciencia contracts: FPA2007-29115-E,
FPA2008-00592 and FIS2006-03438,
U.Complutense/Banco Santander grant PR34/07-15875-BSCH and
UCM-BSCH GR58/08 910309 and the European Community-Research Infrastructure
Integrating Activity
``Study of Strongly Interacting Matter''
(HadronPhysics2, Grant Agreement
n. 227431)
under the Seventh Framework Programme of EU.
\vspace*{-.5cm}

%%%%%%%%%%%%%%%%%%%%%%%%%%%%%%%%%%%%%%%%%%%
%% The following lines show an example how to produce a bibliography
%% without the help of the BibTeX program. This could be used instead
%% of the above.
%%%%%%%%%%%%%%%%%%%%%%%%%%%%%%%%%%%%%%%%%%%

\end{document}